\documentstyle[epsfig,aps,multicol,pre,tighten,amssymb]{revtex}
\begin{document} 
\draft 
\title{Analytical results for coupled map lattices  
with long-range interactions}                
 
\author{Celia Anteneodo$^{1}$, Sandro E. de S. Pinto$^{2}$,  
Ant\^onio M. Batista$^{3}$ and Ricardo L. Viana$^{2}$} 
   
\address{ 
1. Centro Brasileiro de Pesquisas F\'{\i}sicas, Rua Dr. Xavier Sigaud 150,  
22290-180, Rio de Janeiro, RJ, \\ 
2. Departamento de F\'{\i}sica, Universidade Federal do Paran\'a,  
81531-990, Curitiba, PR,\\  
3. Departamento de Matem\'atica e Estat\'{\i}stica, Universidade  
Estadual de Ponta Grossa, Ponta Grossa, PR, Brazil.} 
 
\date{\today} 
\maketitle 
 
\begin{abstract} 
We obtain exact analytical results for lattices of maps   
with couplings that decay with distance as $r^{-\alpha}$.  
We analyze the effect of the coupling range on the system dynamics 
through the Lyapunov spectrum. 
For lattices whose elements are piecewise linear maps,  
we get an algebraic expression for the Lyapunov spectrum.  
When the local dynamics is given by a nonlinear map,  
the Lyapunov spectrum for a completely synchronized state 
is analytically obtained.  
The critical lines characterizing the synchronization transition  
are determined from the expression for the largest 
transversal Lyapunov exponent.  
In particular, it is shown that in the thermodynamical limit,  
such transition is only possible for sufficiently long-range  
interactions, namely, for $\alpha \le \alpha_c<d$, where $d$ is the lattice  
dimension.  

\end{abstract} 
 
 
 
\begin{multicols}{2}
\narrowtext


Synchronization between coupled chaotic systems is one of the most 
intriguing nonlinear phenomena \cite{pecorareview}. 
It has attracted much interest since two decades ago \cite{synchro} 
as it appears in a wide range of real systems   
such as in arrays of Josephson junctions \cite{jjarray}, 
oscillating chemical reactions \cite{kuramotobook}, 
physiological processes \cite{glass}, and has applications as in 
communications \cite{comm} and control theory \cite{control}.
There are many types of synchronized behavior \cite{boccaletti}, but we are 
particularly interested in the completely synchronized states (CSSs) of coupled 
map lattices (CMLs), where all maps present the same amplitude at all times. 
Complete synchronization is an example of non-equilibrium phase 
transition \cite{ap02}, which may be related to actual critical phenomena 
like the superconducting-normal transition in Josephson junctions \cite{supercond}.

CMLs, which are dynamical systems with discrete space and time, 
and a continuous state variable, have been investigated as theoretical 
models of spatiotemporal phenomena in a variety of 
problems in condensed matter physics, neuroscience and chemical physics 
\cite{kanekobook}. 
The spatiotemporal behavior is governed by two simultaneous mechanisms: 
the intrinsic nonlinear dynamics of each map, and diffusion due to the 
spatial coupling between maps; the dynamical pattern being the outcome 
of the competition between them. This applies, in particular, 
to the problem of synchronization of chaotic maps \cite{synchaoscml}. 
The effective coupling range is a crucial factor to determine whether 
or not chaotic maps mutually synchronize. Nearest-neighbor couplings 
(short range) do not favor synchronization, since the coupling effect 
is typically too weak to overcome the intrinsic randomness of map dynamics 
\cite{kaneko1}. On the other hand, long-range couplings tend 
to facilitate synchronization, as exemplified by the limiting case of global 
(mean-field) coupling \cite{kaneko3}. Lattices of non-locally coupled maps 
appear in neural networks with local production of information \cite{nozawa}, 
models of physico-chemical reactions \cite{gade1}, assemblies of biological 
cells with oscillatory activity \cite{kuramoto}, and diffusion coupling in 
nucleation kinetics \cite{batok}. 
Beyond CMLs, systems with many degrees of freedom with long-range couplings  
are an interesting object of study because of their anomalies 
(appearing at the level of the macroscopic thermodynamical description as well 
as in the underlying microscopic dynamics), which still require deeper 
understanding \cite{lrange}. 
Simple dynamical models, such as CMLs, may add new knowledge on 
non-equilibrium long range systems. 
However, there is a lack of analytical results for CMLs with 
arbitrary range couplings.  
Exact analytical results are particularly crucial because 
the occurrence of phenomena such as shadowing breakdown 
\cite{shadowbreak} or spurious synchronization \cite{spurioussync}  
set difficulties in numerical approaches due to the unavoidable 
finite precision of numerical simulations. 

Here we examine a form of coupling whose 
intensity decays with the distance $r$ between sites as $1/r^{\alpha}$, 
with $\alpha\geq 0$ \cite{sandro1}. 
It has also been considered  in biological networks \cite{ragha}, 
in ferromagnetic spin models \cite{spins},  
many-particle conservative (Hamiltonian time evolution) classical  
systems \cite{celia1,celia2}, large populations of limit cycle  
oscillators \cite{monica} and a generalization of the  
Kuramoto model \cite{rogers}, among other examples.  
Explicitly, we consider a chain of $N$ coupled one-dimensional chaotic 
maps $x \mapsto f(x)$ such that the coupling  prescription is  
 
\begin{equation} 
x^{(i)}_{n+1}=(1-\varepsilon)f(x^{(i)}_{n})+\frac{\varepsilon} 
{\eta(\alpha)}\sum^{N'}_{r=1} 
\frac{f(x^{(i-r)}_{n})+f(x^{(i+r)}_{n})}{r^{\alpha}}, 
\label{CML} 
\end{equation} 
where $ x^{(i)}_n$ represents the state variable for the site $i$ $(i=1,2,...,N)$  
at time $n$, $\varepsilon \geq 0$ and $\alpha \geq 0$ are the coupling  
strength and effective range,  
respectively, and $\eta(\alpha)=2 \sum^{N'}_{r=1}r^{-\alpha}$, is the normalization  
factor, with $N'=(N-1)/2$ for odd $N$.  
In conservative systems \cite{celia1,celia2}, scaling by $\eta$ plays an important  
role in  making the systems pseudo-extensive.   
Here periodic boundary conditions  
$x^{(i)}_{n}=x^{(i\pm N)}_n$ and random initial conditions are assumed.   
The coupling term is a weighted average of discretized spatial second derivatives,  
the normalization factors being the sum of the corresponding statistical weights.  
It is straightforward to prove that in the limits $\alpha=0$ and  
$\alpha \rightarrow\infty$ Eq. (\ref{CML}) reduces to the global mean-field  
and the local Laplacian-type couplings, respectively.  
 
We characterize the spatio-temporal synchronization dynamics by means of the 
Lyapunov spectrum (LS) of the lattice, that enables one to estimate, for 
instance, the Kolmogorov-Sinai entropy through the Pesin formula \cite{isola} and 
the Lyapunov dimension, which gives an upper bound on the effective 
number of degrees of freedom needed to characterize the system dynamics 
\cite{ruelle}. 
Besides characterizing a CSS, when it exists at all, 
we must investigate its stability with respect to small perturbations. 
If the CSS turns out to be dynamically unstable, we are faced with two 
possibilities: either the CSS presents the so called bubbling attractor, 
and in this case the CSS only lasts for a finite time, or the CSS 
loses transversal stability through a blowout bifurcation \cite{boccaletti}. 

In this work we will present exact analytical results for the CML (\ref{CML}). 
We will show that for a 1D lattice of $N$ coupled piecewise linear maps  
it is possible to obtain an exact analytical expression  
for the LS, the results shown in  
\cite{kaneko3,isola} being recovered in the limits  
$\alpha=0$ and $\alpha \rightarrow \infty$.  
When the maps $x\mapsto f(x)$ are nonlinear,  
we will show that analytical results are still possible  
for CSSs.   
By means of the algebraic formulas for the LS, one can find   
the synchronization regions in the $\varepsilon \times \alpha$ space,  
since the second largest Lyapunov exponent (belonging to the direction 
transversal to the SM) equal to zero   
indicates a transition to the synchronized state \cite{gade1,gade2}. 
Finally, the results obtained for a chain of maps 
will be extended to $d$-dimensional hypercubic lattices.  

 
In order to calculate the LS one has to consider the tangent dynamics.  
By differentiating the equations of the original maps (\ref{CML}),  
one obtains the evolution equations for tangent vectors  
$\xi =(\delta x^{(1)},\delta x^{(2)},\ldots,\delta x^{(N)})^T$, that in  
matrix form read  $ \xi_{n+1}={\bf T}_n\xi_{n}$, 
with the Jacobian matrix ${\bf T}_n$ given by 
 
\begin{equation} 
{\bf T}_n=\left[ (1-\varepsilon)+\frac{\varepsilon} 
{\eta(\alpha)}{\bf B}\right]{\bf D}_n, 
\label{tanmatrix} 
\end{equation} 
where the matrices ${\bf D}_n$ and ${\bf B}$ are defined, respectively, by 
$D_{n}^{jk}=  f^\prime(x^{(j)}_n) \delta_{jk}$ and  
$ B_{jk}= 1/r^{\alpha}_{jk} (1-\delta_{jk})$ , 
being $r_{jk}=\mbox{min}_{l\in \cal{Z}}|j-k+lN|$. Notice that the  
particular choice of the interaction law is embodied in the matrix ${\bf B}$  
which is time independent. 
 
Once specified the initial conditions, the LS is extracted  
from the evolution of the initial tangent vector $\xi_0$:  
$ \xi_n = {\cal T}_n\xi_0$, 
where ${\cal T}_n\equiv {\bf T}_{n-1}\dots {\bf T}_{1}{\bf T}_{0}$  
is product of $n$ Jacobian matrices calculated at successive points of the discrete   
trajectory. If $\Lambda_{1},\ldots ,\Lambda_{N}$ are the eigenvalues of  
$\hat{\Lambda}= {\displaystyle \lim_{n\rightarrow \infty}} 
( {\cal T}^{T}_n {\cal T}_n)^{\frac{1}{2n}} $  
(that are real and positive), the Lyapunov exponents  
are obtained as \cite{eckmann} 
 
\begin{equation}  \label{lyap1}  
\lambda_k=\ln\Lambda_k, \;\;\;\;\;\;\;\; k=1,\ldots,N. 
\end{equation} 
 

We start by applying the expressions above to 
the piecewise linear maps $x\mapsto f(x)=\beta x$ (mod 1),  
with $\beta\geq 1$. In this case we have $f^{\prime}(x)=\beta=\mbox{constant}$,  
therefore ${\bf D}_n=\beta \openone_{N}$, and ${\bf T}_n$ becomes 
 
\begin{equation} 
{\bf T}_n=\beta\left[(1-\varepsilon)\openone_{N}+ 
\frac{\varepsilon}{\eta(\alpha)}{\bf B}\right]\equiv \beta \hat{\bf B}\, , 
\label{tanmatrix1} 
\end{equation} 
the rightmost identity defining the matrix $\hat{\bf B}$.  
Since the symmetric tangent map does not depend on time, 
it results $\hat\Lambda=\beta\hat{\bf B}$. So, in order to obtain the LS,  
it is enough to diagonalize ${\bf B}$. 
Because of its periodicity, ${\bf B}$ can be diagonalized  
in Fourier space \cite{eigen}, the eigenvalues being 
 
\begin{equation} 
b_{k}=2\sum^{N^{\prime}}_{m=1}\frac{\cos(2\pi km/N)}{m^{\alpha}}, 
\;\;\;\;\;\;\;\; k=1,\ldots,N, 
\label{beigenvalues} 
\end{equation} 
where we considered odd $N$. 
Finally, from Eq. (\ref{lyap1}), taking into account the special form of  
$\hat{\Lambda}$, the LS is given by  
 
\begin{equation} 
\lambda_{k} \,=\, 
\ln\beta +\ln \left\vert 
1-\varepsilon +\frac{\varepsilon}{\eta(\alpha)}b_k\right\vert  \, .   
\label{spectrum} 
\end{equation} 
This expression is consistent with previous numerical results \cite{sandro2}.  
In the extreme cases  $\alpha\rightarrow\infty$ and $\alpha=0$ the known 
expressions  \cite{kaneko3,isola} are recovered.  
 

Now we will consider lattices of nonlinear maps.   
An important case that can be tackled easily is the one where the maps are  
in the CSS.  
As it will become clear soon, this instance provides relevant information 
on the synchronization transition. 
In the CSS, the dynamical variables of all maps coincide, i.e, 
$x^{(1)}_n=x^{(2)}_n=\ldots, x^{(N)}_n\equiv x^{(*)}_n$, at each time step $n$.   
The LS for the CCS when $\alpha=0$ has already been found  
by Kaneko \cite{kaneko3}. Now, for arbitrary $\alpha$, we have  
${\bf D}_n=f^\prime(x^{(*)}_n)\openone_{N}$, thus, 
${\bf T}_n=f^\prime(x^{(*)}_n)\hat{\bf B}$  
and ${\cal T}^{T}_n {\cal T}_n= (\prod_{j=0}^{n-1} 
[f^\prime(x^{(*)}_j)]^{2}) \hat{\bf B}^{2n}$.  
Therefore, for the CSS, following Eq. (\ref{lyap1}), one arrives at 
 
\begin{equation} \label{beigenstar}
\lambda^*_k\,=\,\lim_{n\rightarrow\infty}
\frac{1}{n}\sum^n_{i=1}\ln|f^\prime(x^{(*)}_i)| 
+\ln|1-\varepsilon +\frac{\varepsilon}{\eta(\alpha)}b_k|, 
\end{equation} 
where $b_k$ are the eigenvalues of ${\bf B}$ defined in (\ref{beigenvalues}). 
Assuming ergodicity, the time-average in (\ref{beigenstar}) 
can be substituted by an average over the single-map attractor.  
In this way one gets 
 
\begin{equation} 
\lambda^*_k\,=\,\lambda_{U} + \ln
\left\vert 1-\varepsilon +\frac{\varepsilon}{\eta(\alpha)}b_k \right\vert,  
\label{spectrum_CSS} 
\end{equation}
where $\lambda_{U}=\langle \ln|f^\prime(x^{(*)})| \rangle$ 
is the Lyapunov exponent of an uncoupled map. 
This expression is {\em general}: it applies to any lattice of 
nonlinear 1D maps coupled 
with the scheme here considered, the parameters that define the particular  
uncoupled map affecting only $\lambda_U$. 
For instance, for the logistic map $x\mapsto f(x)=ax(1-x)$,  
with $a=4$ and $x \in [0,1]$,  
$\lambda_{U}=\langle \ln |4(1-2x^{(*)})|\rangle=\ln 2$ \cite{ott}  
and the contribution of the power-law coupling is always  
$\ln|1-\varepsilon +\frac{\varepsilon}{\eta(\alpha)}b_k|$. 
Notice that the LS in CSSs has the same structure as 
the LS obtained for piecewise linear maps 
[Eqs. (\ref{beigenvalues})-(\ref{spectrum})]. 
 

The synchronization transition can be characterized by a complex 
order parameter \cite{kuramotobook} defined, for time $n$, as 
$R_n=| \frac{1}{N} \sum_{j=1}^{N} e^{2\pi i x_n^{(j)}}|$. 
A time-averaged amplitude $\bar R$ is computed over an interval 
large enough to warrant that the lattice has attained the asymptotic state.  
In the CSS, one has $\bar R =1$. 
On the opposite case of completely non-synchronized maps, 
the site state variables $x_n^{(j)}$ are so 
uncorrelated that $\bar R \approx 0$. 
 
A diagnostic of synchronization can also be extracted from the LS. 
It can be easily verified that, for arbitrary $\alpha$,  
the CSS lies along  
the direction of the eigenvector associated to the largest exponent.
This was previously observed by Kaneko for the particular case 
$\alpha=0$ \cite{kaneko3}.  
Therefore, the CSS will be transversally 
stable if the $(N-1)$ remaining exponents are non positive,  
that is, $\widetilde\lambda^*_2< 0$  
(where the tilde stands for ordered exponents). 
The second largest exponent, $\widetilde\lambda^*_2$, corresponds in Eq. 
(\ref{spectrum_CSS}),  to $k=1$ (or $k=N-1$, due to degeneracy) 
if the argument of the modulus in Eq. (\ref{spectrum_CSS}) is positive 
and to $k=(N-1)/2$ (or $k=(N+1)/2$) otherwise. 
One obtains  
 
\begin{eqnarray}  \label{critical}
\varepsilon_c&=& 
 ( 1-{\rm e}^{-\lambda_U} ) 
 \biggl(1- \frac{2}{\eta(\alpha)} \sum_{m=1}^{N^\prime} 
\frac{\cos(2\pi m/N)}{m^{\alpha}}\biggr)^{-1}, \\ 
\varepsilon^\prime_c&=& 
 ( 1+{\rm e}^{-\lambda_U} ) 
 \biggl(1- \frac{2}{\eta(\alpha)} \sum_{m=1}^{N^\prime} 
\frac{\cos(\pi m(N-1)/N)}{m^{\alpha}}\biggr)^{-1}, \nonumber
\end{eqnarray} 
where $\varepsilon_c$ ($\varepsilon^\prime_c$ ) are the coupling strengths below (above) 
which the SM ceases to be transversely stable, such that synchronized states are 
not typically observed. 
It is noteworthy that Eq. (\ref{critical}) is {\em quite general} for  
the coupling scheme here considered:  the parameters that define the 
particular uncoupled nonlinear map (embedded in $\lambda_U$) just 
participate through the first factor.

\begin{figure}[htb] 
\begin{center} 
  \includegraphics*[bb=40 150 550 650, width=0.4\textwidth]{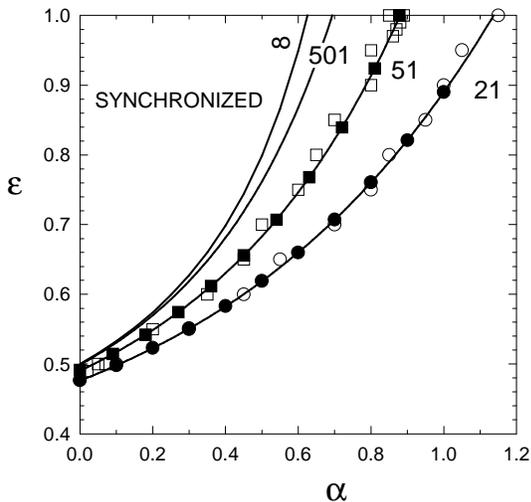}
\end{center} 
\caption{\protect Synchronization in parameter plane  
$\varepsilon\times\alpha$ for a 1D lattice of $N$ coupled logistic maps   
$x\mapsto f(x)=4x(1-x)$  and $\varepsilon \leq 1$.  
The critical line $\varepsilon_c(\alpha)$ was determined  analytically 
from Eq. (\ref{critical})  
for different values of $N$ (full lines) and numerically from the conditions  
$\bar R \neq 1$ (open symbols) and $\tilde{\lambda_2^*} =0$ (full symbols) 
for $N=21$ and 51.} 
\end{figure}

Fig.~1 presents {the lower critical line} in the parameter space  
$\varepsilon \times \alpha$ for $N$ coupled logistic maps   
$x\mapsto f(x)=4x(1-x)$.  
The critical line was obtained analytically from Eq. (\ref{critical})  
and numerically by means of two procedures: either  
when $\bar R\neq 1$ is numerically detected  
or by the condition of nullity 
of the second largest Lyapunov exponent in the CSS.  
The numerical results are in good agreement with the analytical prediction. 

As can be observed in Fig. 1, the critical frontier depends on the system size $N$. 
In the limit $N\to\infty$ we obtain 
 
\begin{equation} 
\varepsilon_{c,\infty} \;=\;  
\frac{1- {\rm e}^{-\lambda_U} }{1-C(\alpha)}, 
\label{critical_inf} 
\end{equation} 
where $C(\alpha)$ corresponds to 
 
\begin{equation} \label{calfa}
C(\alpha)\;=\;  \lim_{N\to\infty} 
 \biggl( \biggl[ \sum_{m=1}^{N^\prime} \frac{\cos(2\pi m/N)}{m^{\alpha}} \biggr]  
         \biggl[ \sum_{m=1}^{N^\prime} \frac{1}{m^{\alpha}} \biggr]^{-1}  \biggr).  
\end{equation} 
This limit is equal to unity for $\alpha >1$,   
so that Eq. (\ref{critical_inf}) furnishes a divergent result. For  
$\alpha$ outside the domain of convergence of the series, one gets 
 
\begin{equation} 
C(\alpha)\;=\;   
\frac{1-\alpha}{\pi^{1-\alpha}} \int_0^\pi \frac{\cos(x)}{x^\alpha}\,dx . 
\end{equation} 
 
Moreover, one has $\varepsilon^\prime_{c,\infty} \;=\;  
1+ {\rm e}^{-\lambda_U}$, if $\alpha<1$. 
Then, in the  limit $N\to\infty$, synchronization  
is only possible for sufficiently long-range interactions (see Fig. 2), 
namely, for $\alpha \leq \alpha_c <1$.  
The  critical value here obtained for 1D CMLs    
is different from the one reported for  other 
1D systems with similar power-law interactions, such as   
ferromagnetic spin models \cite{spins} or  
many-particle classical Hamiltonian systems \cite{celia1};  
in such cases, the critical value for the existence of an  
order/disorder transition is $\alpha_c=2$. 
In a generalization of the Kuramoto model, the value of $\alpha_c$ is controversial:  
While $\alpha_c=2$  was first reported \cite{rogers}, recent analytical considerations 
point to $\alpha_c=1$ \cite{alfacrit}.   
Although the  generalized Kuramoto model is a continuous time dynamical system, 
our analytical result for the upper bound of the critical value suggests that the latter value 
is the correct one. 

The expressions we have derived for the LS of chains of maps 
can be straightforwardly generalized 
for hypercubic lattices of arbitrary dimension $d$. In fact, for the general 
$d$-dimensional case, it is straightforward to show that 
the eigenvalues of the matrix ${\bf B}$ become 

\begin{equation} 
b_{k} \,=\, 2^d \sum_{\bar{m}\neq0} 
\frac{\cos ( 2\pi \bar{r}_k.\bar{m}/N^\frac{1}{d} )}{m^\alpha} \, ,
\;\;\;\;\;\;\;\; k=1,\ldots,N, 
\label{beigenvalues_d} 
\end{equation} 
where $\bar{r}_k$ is the position vector of site $k$, 
$\bar{m}=(m_1,\ldots,m_d)$, with $0\leq m_i\leq N^\prime$, 
and $N^\prime=(N^\frac{1}{d}-1)/2$. The normalization factor reads
\begin{equation} 
\eta(\alpha,d) \;=\; 2^d \sum_{\bar{m}\neq0} 
\frac{1}{m^\alpha} 
\;\;\propto\;\frac{N^{\alpha/d}-1}{1-\alpha/d}\, .
\label{eta_d} 
\end{equation} 

Then, following the same lines as before but now for the sake of 
generalizing Eq. (\ref{critical_inf}), it  is easy to see 
that $\varepsilon_{c,\infty}$ will diverge if $\alpha/d>1$, which leads to $\alpha_c < d$.

Additionally, our results for the LS could be extended to the more general class of 
coupling schemes where the dependence of the coupling strength on the inter-map 
distance is not necessarily of the power-law type.  
In these cases, one should feed Eqs. (\ref{spectrum}) and  (\ref{spectrum_CSS}) 
with the eigenvalues of the appropriate matrix ${\bf B}$, that
contains the particular dependence of the interaction strength on distance.

\newpage
\begin{figure}[htb] 
\begin{center} 
 \includegraphics*[bb=40 170 550 650, width=0.4\textwidth]{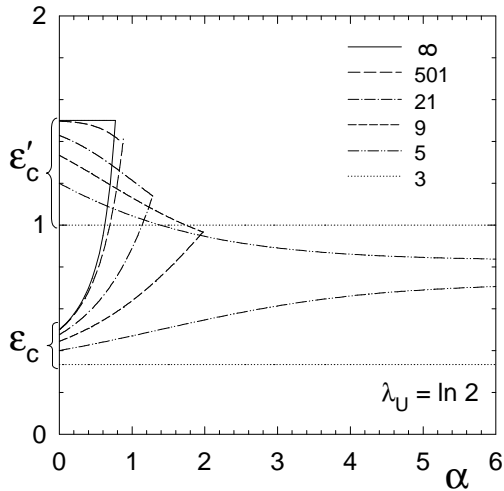}
\end{center} 
\caption{\protect 
Critical coupling strengths for synchronization 
vs range parameter, for different values of $N$ indicated in the figure. 
The critical lines were analytically obtained from Eq. (\ref{critical})  
for a 1D lattice with power-law couplings and $\lambda_U =\ln 2$. 
Synchronization occurs for $\varepsilon_c<\varepsilon <\varepsilon^\prime_c$.
} 
\end{figure}


In conclusion, we have presented analytical expressions for the LS
of CMLs with an interaction which decays with the
lattice distance as a power law, for two cases: (i) 
piecewise linear coupled maps; and (ii) the CSS of lattices of 
one-dimensional maps. Our results enable us to predict the
critical values for synchronization in the coupling parameter plane
(strength {\it versus} range), and also may be used to obtain related
quantities of interest, like KS-entropies and Lyapunov dimensions. 
Such exact analytical results are crucial in order to avoid difficulties
present in numerical approaches, such as shadowing breakdown,  
due to unavoidable finite precision of numerical simulations. 
In addition, we have shown that, in the thermodynamical limit, the critical
range for synchronization is equal to the lattice dimension. Many of our
results could be extended to lattices of continuous-time oscillators,
and hence have an even wider range of applicability.

\section*{Acknowledgments:}
We are grateful to S.R. Lopes, C. Tsallis and  R.O. Vallejos for useful comments, 
and  to S. Ruffo for old fruitful remarks. 
This work was partially supported by brazilian agencies CAPES, CNPq, FAPERJ,  
Funda\c{c}\~ao Arauc\'aria e PRONEX.

\end{multicols}
\end{document}